\documentclass{sigchi}
\pdfoutput=1

\toappear{}

\usepackage{balance}  
\usepackage{txfonts}
\pdfmapfile{+txfonts.map}

\usepackage{times}    
\usepackage{color}
\usepackage{textcomp}
\usepackage{ccicons}
\usepackage{todonotes}

\usepackage{graphicx}
\usepackage[pdftex]{hyperref}
\usepackage{booktabs}
\usepackage{multirow}
\usepackage{multicol}
\usepackage{balance}
\usepackage{enumitem}
\usepackage[super]{nth}

\usepackage{url}

\newcommand*\rot{\rotatebox{90}}

\makeatletter
\def\url@leostyle{%
  \@ifundefined{selectfont}{\def\UrlFont{\sf}}{\def\UrlFont{\small\bf\ttfamily}}}
\makeatother
\def\pprw{8.5in}
\def\pprh{11in}

\definecolor{linkColor}{RGB}{6,125,233}
\hypersetup{%
  pdftitle={SIGCHI Conference Proceedings Format},
  pdfauthor={LaTeX},
  pdfkeywords={SIGCHI, proceedings, archival format},
  bookmarksnumbered,
  pdfstartview={FitH},
  colorlinks,
  citecolor=black,
  filecolor=black,
  linkcolor=black,
  urlcolor=linkColor,
  breaklinks=true,
}

\begin{document}

\title{Why Johnny Still, Still Can't Encrypt:\\Evaluating the Usability of a Modern PGP Client}

\numberofauthors{1}
\author{
\alignauthor
Scott Ruoti, Jeff Andersen, Daniel Zappala, Kent Seamons\\
	\affaddr{Brigham Young University}\\
	\email{\{ruoti, andersen\} @ isrl.byu.edu, \{zappala, seamons\} @ cs.byu.edu}
}

\maketitle

\begin{abstract}
This paper presents the results of a laboratory study involving Mailvelope, a modern PGP client that integrates tightly with existing webmail providers.
In our study, we brought in pairs of participants and had them attempt to use Mailvelope to communicate with each other.
Our results shown that more than a decade and a half after \textit{Why Johnny Can't Encrypt}, modern PGP tools are still unusable for the masses.
We finish with a discussion of pain points encountered using Mailvelope, and discuss what might be done to address them in future PGP systems.
\end{abstract}

\keywords{Security, usability, secure email, PGP}

\category{H.1.2.}{Models and Principles}{User/Machine Systems}[human factors]
\category{H.5.2.}{Information Interfaces and Presentation (e.g. HCI)}{User Interfaces}[user-centered design]

\section{Introduction}
Usable, secure email is still an open problem more than 15 years after it was first studied by Whitten et al.~\cite{whitten1999why}.
Six years after the original Johnny paper, Sheng et al. showed that PGP 9 was still difficult for users to operate correctly~\cite{sheng2006why}.
In this paper, we attempt to see if in the last decade, modern PGP-based tools have improved to the point where users can successfully send encrypted email.

We elected to test Mailvelope, a modern PGP tool, for our study.
Mailvelope is a browser extension that integrates with users' webmail systems.
It is the only system currently being promoted by the EFF's secure message score card\footnote{\url{https://www.eff.org/secure-messaging-scorecard}} that integrates with users' webmail providers, an important feature for many users.
It is also highly rated on the Chrome Web Store, with 242 users collectively giving it 4.6 out of 5 stars.
In our own testing of PGP alternatives, we found Mailvelope to be roughly as usable as other alternatives (i.e., GPG Tools, Enigmail, Google's End-to-End Encryption).

In our study of 20 participants, grouped into 10 pairs of participants who attempted to exchange encrypted email, only one pair was able to successfully complete the assigned tasks using Mailvelope.
All other participants were unable to complete the assigned task in the one hour allotted to the study.
This demonstrates that encrypting email with PGP, as implemented in Mailvelope, is still unusable for the masses.

Our results also shed light on several ways that PGP-based tools could be improved.
First, integrated tutorials would be helpful in assisting first time users in knowing what they should be doing at any given point in time.
Second, an approachable description of public key cryptography could help users correctly manage their own keys.
Third, in line with previous work by Atwater et al. \cite{atwater2015leading}, we find that PGP-based tools would be well served by offering automatically generated emails for unknown recipients asking them to install the PGP software, generate a public key, and share it with the sender.
Finally, the PGP block itself could be enhanced to help non-PGP users who receive an encrypted email know how to work with their friend to get an encrypted message they will be able to read.

\section{Related Work}

Whitten and Tygar~\cite{whitten1999why} conducted the first formal user study of a secure email system (i.e., PGP 5), uncovering serious usability issues with key management and users' understanding of the underlying public key cryptography. It was found that a majority of users were unable to successfully send encrypted email in the context of a hypothetical political campaign scenario. The results of the study took the security community by surprise and became responsible for shaping modern usable security research.
Sheng at al. demonstrated that despite improvements made to PGP in the seven years since Whitten and Tygar's original publication, key management was still a challenge for users. Furthermore, they showed that in the new version of PGP encryption and decryption had become so transparent that users were unsure if a message they received had actually been encrypted.

While there have been some attempts at implementing email using key escrow \cite{fahl2012helping,ruoti2013confused}, these approaches have degraded security when compared to traditional PGP.
Atwater et al. created a mock-up of Mailvelope that automated key creation and sharing.
This approach falls somewhere in between traditional PGP and key escrow on the spectrum of usability and security tradeoffs.
Unfortunately, it was not possible to include Atwater et al.'s mock-up in our study, as we discovered that it relied on hard-coded keys for email recipients, and the effort required to implement a working key management system into their mock-up would have exceeded our resources.
Furthermore, the mock-up did not correctly simulate the need for participants to wait on the recipient to set up keys, which made the mock-up incompatible with our study.\footnote{This incorrect simulation also calls into question their results regarding the high usability of automated-PGP.}

\section{Methodology}
We conducted an IRB-approved user study wherein pairs of participants used secure email to transmit sensitive information to each other.
This section gives an overview of the study and describes the scenario, tasks, study questionnaire, and post-study interview. In addition, we discuss how the study was developed and its limitations.

\subsection{Study Setup}
The study ran for two weeks, beginning Tuesday, September 8, 2015 and ending Friday, September 18, 2015.
In total, 10 pairs of participants (20 total participants) completed the study.
Participants were allocated sixty minutes to complete the study, with about 35-40 minutes spent using Mailvelope.
Participants were compensated \$15 USD for their participation.
Participants were required to be accompanied by a friend, who served as their counterpart for the study.
For standardization and requirements of the systems tested in the study, both participants were required to have Gmail accounts.

When participants arrived, they were read a brief introduction detailing the study and their rights as participants.
Participants were informed that they would be in in separate rooms during the study and would use email to communicate with each other.\footnote{The study coordinators ensured that the participants knew each other's email addresses.}
Participants were also informed that a study coordinator would be with them at all times, and could answer questions they might have.

\subsection{Demographics}
We recruited Gmail users for our study at a local university.
Participants were two-thirds male: male (13; 65\%), female (7; 35\%).
Participants skewed young: 18 -- 24 years old (18; 90\%), 25 -- 34 years old (2; 10\%).

We distributed posters broadly across campus to avoid biasing our results to any particular major.
All participants were university students,\footnote{We did not require this.} with the majority being undergraduate students: undergraduate students (17; 85\%), graduate students (3; 15\%).

\subsection{Scenario and Task Design}
During the study, participants were asked to role-play a scenario regarding completing taxes. Participant A was told they needed Participant B's help with filing taxes. Participant A was also told they since they were sending sensitive information (e.g., SSN) that they should encrypt this information using Mailvelope.\footnote{Participants used sensitive information we generated and not their own information.} Participant B was told to wait for his friend to send him the necessary sensitive information (e.g., SSN). Once Participant B had received this information, he was instructed to use Mailvelope to respond to Participant A with a confirmation code (encrypted using Mailvelope) to conclude the task.

After the instructions were given, Participant A was provided with the the Mailvelope website and instructed to begin the task.\footnote{\url{https://www.mailvelope.com/}}
While participants waited for email from each other, they were told that they could browse the Internet, use their phones, or engage in other similar activities.
This was done to provide a more natural setting for the participants, as well as to avoid frustration if participants had to wait for an extended period of time while their friends figured out how to use Mailvelope.\footnote{In some cases Participant B never actually received an email from Participant A.}

Study coordinators were allowed to answer questions related to the study, but were not allowed to provide instructions on how to use Mailvelope.
If participants became stuck and asked for help, they were told that they could take whatever steps they normally would to solve a similar problem, including using an Internet search service. Additionally, when asked for help, if the study coordinator believed communication between the two parties could help, study coordinators could remind participants that they were completing this study with their friend and were free to communicate with their friend however they wanted, and that only the sensitive information was required to be transmitted over secure email.

\subsection{Study Questionnaire}
We administered our study using the Qualtrics web-based survey software.
Before beginning the survey, participants were read an introduction by the study coordinator and then asked to answer a set of demographic questions.
Participants then completed the study task.

Immediately upon completing the study task, participants were asked several questions related to their experience with that system.
First, participants completed the ten System Usability System questions \cite{brooke1996sus,from2013sus}.
Second, participants were also asked to describe in their own words what they liked about Mailvelope, what they would change, and why they would change it.
If Participant B had never received an encrypted email from Participant A, they were not required to complete the study questionnaire, as they had no experience with Mailvelope.

\subsection{Post-study Interview}
After the completion of the survey, participants were interviewed by their respective study coordinators.
Study coordinators focused on issues that had arisen during the study and probed for more details regarding areas of confusion.
After the participants completed their individual post-study interviews, they were brought together for a final post-study interview.
Participants were asked to explain what they had experienced to each other, and notes were taken on what was said by the study coordinators.
Additionally, participants were asked how an ideal secure email system would function.
While participants are not system designers, our experience has shown that when asked to design ideal systems, participants often reveal preferences that otherwise remain unspoken.

\subsection{Survey Development and Limitations}
After developing the study, we conducted a pilot study with three pairs of participants (six participants total).
Based on the results of the pilot study, we printed out the task instructions and provided them to the participants to reference while completing the study.

Our study only included twenty participants, all of whom were students. While this was enough to show difficulties associated with Mailveope it is not indicative of all possible outcomes.
It would be especially interesting to rerun this study using different populations (e.g., technical professionals, computer scientists, security professionals).
Future studies could also be conducted to see how Mailvelope fares when some of its more glaring problems are addressed.

\section{Results}

In this section, we report on the SUS scores for Mailvelope, provide details on the success rate, and list the mistakes made by participants.

\subsection{System Usability Scale}

\begin{table}
\centering
\resizebox{\columnwidth}{!}{

\begin{tabular}{l|c|cc|ccccc|}

	\rule{0pt}{11ex} & \rot{Count} & \rot{Mean} & \rot{\shortstack[1]{Standard\\Deviation}} & \rot{Min} & \rot{Q1} & \rot{Median} & \rot{Q3} & \rot{Max} \\
	\midrule
	
	Participant A 	& 10 & 30.5 & 16.6 & 0.0  & 21.3 & 28.8 & 41.3 & 57.5 \\ 
	Participant B	& 6  & 41.3 & 10.9 & 27.5 & 33.75 & 41.3 & 16.9 & 57.5 \\ 
	Combined		& 16 & 34.5 & 15.3 & 0.0  & 25.0 & 35.0 & 45.6 & 57.5 \\ 
		
	\bottomrule
\end{tabular}
}

\caption{SUS Scores}
\label{tab:sus}
\end{table}

We evaluated Mailvelope using the System Usability Scale (SUS).
A breakdown of the SUS score for each system and type of participant (i.e., Participant A, Participant B, or combined) is given in Table~\ref{tab:sus}.
The mean value is used as the SUS score \cite{brooke1996sus}.

\begin{figure*}[t]
\centering
\includegraphics[width=1.0\textwidth]{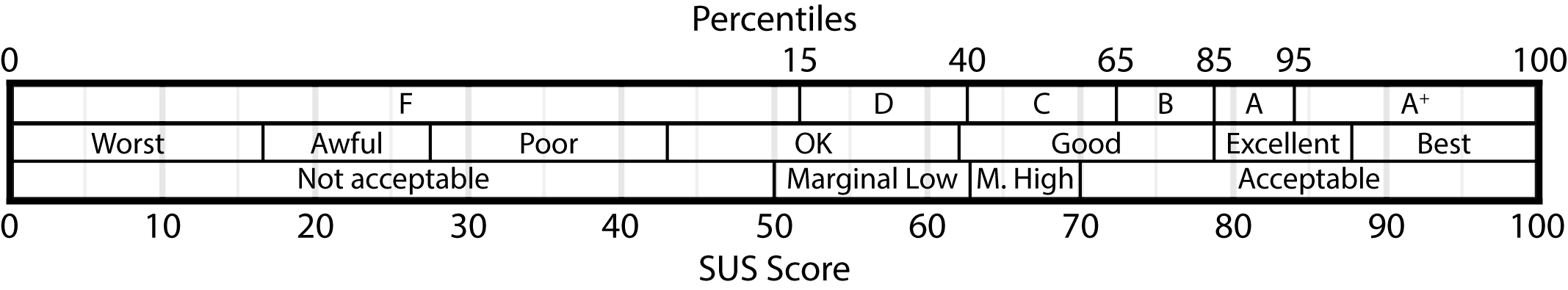}
\caption{Adjective-based Ratings to Help Interpret SUS Scores}
\label{fig:sus}
\end{figure*}

To give greater context to the meaning of each system's SUS score, we leverage the work of several researchers who analyzed hundreds of studies which used SUS in order to give adjective-based ratings describing SUS scores~\cite{bangor2008empirical,bangor2009determining,sauro2011practical}.
Based on this data, we generated ranges for these adjective ratings, such that a score is correlated with the adjective it is closest to in terms of standard deviations (see Figure~\ref{fig:sus}).
Using these ratings, Mailvelope's SUS score of 34.5 is rated as having ``Poor'' usability.
It falls below the \nth{15} percentile, is given a letter grade of ``F'' and is labeled as ``Not acceptable''.

\subsection{Failures}
Participants were given between 30 and 45 minutes to complete the tasks.
Study coordinators were instructed to end the task if thirty minutes elapsed after Participant A had first installed Mailvelope. If participants were making progress, study coordinators would allow them to go up to ten minutes longer, as long as it did not cause the overall study time to exceed one hour.

Of the ten participant pairs, nine were unable to successfully complete the task.
In two of the nine pairs, participants A never figured out how to use Mailvelope to send any message.
In another two pairs, Participant B was completely mystified by the encrypted PGP email and was unaware that they needed to install Mailvelope to read the message.
Only one of the nine pairs actually traded public keys, though this pair was still confused about what to do after sharing their public keys.

The one pair that did complete the task required the full forty-five minutes to do so.
The successful pair was unique in that they were the only pair of participants where one of the participants had previously learned about public key cryptography.
It is likely that this heavily influenced their ability to finish within the time limit.

\subsection{Mistakes}
All participant pairs made mistakes.
The most common mistake was encrypting a message with the sender's public key. This occurred for seven of the participant pairs, including for the participant pair that was eventually successful.
Three of the participant pairs generated a key pair with their friend information, and then tried to use that public key to encrypt their message.
One participant modified the PGP block after encryption (while still in the PGP compose window), adding their sensitive information to the area before the PGP block.
Finally, one participant eventually exported his private key and sent it along with his keyring password to his friend so that his friend could decrypt the message he had received. In this case, even though the participants had transmitted the required information, they were informed that they needed to try some more and accomplish the task without sending the private key.

\section{Discussion}
In this section we discuss themes we found while running this study and include quotes from participants.
Participants have all been assigned a unique identifier M[1-10][A,B].
The final letter refers to which role the participant played during the study, and participants with the same number were paired with each other (e.g., M1A and M1B were Participant A and Participant B, respectively, in the same study session).

Mailvelope clearly failed to help the majority of participants encrypt their email.
All participants expressed frustration with Mailvelope, with the most comical expression of this frustration coming from M3A: \textit{``Imagine the stupidest software you would ever use, and that was what I was doing.''}.
The difficulty also led several participants to indicate that in the real world they would have given up trying to use Mailvelope long before they did during the study.
For example, M3A also said, \textit{``After five minutes, I would have just given up and called.''}

While it is unclear if PGP will ever be usable by the masses, we spend the remainder of this section reporting lessons learned from the study that could assist the design of PGP-based secure email tools.

\subsection{Integrated Tutorials}
Every single participant constantly flipped back and forth between Gmail and Mailvelope's instructions.
At no stage was it intuitive what they should do next based on the Mailvelope UI.
Nearly all participants indicated that they wished Mailvelope had provided instructions that were integrated with the Mailvelope software, and would walk them through, step-by-step, in setting up Mailvelope and sending their first encrypted email.
This has been shown to greatly improve the usability of other secure email systems~\cite{ruoti2013confused}, and would greatly assist first-time users in acclimating to PGP.

Key steps that could be addressed by tutorials are: (1) helping participants generate their PGP key pair, (2) discussing how to share public keys, (3) inviting their friends to setup Mailvelope, (4) importing their friend's public keys, (5) sending their first encrypted email, and (6) decrypting their first encrypted email.
	
\subsection{Explanation of Public Key Cryptography}
The only participant pair that successfully completed the study task likely did so because one of the participants in the pair had previous knowledge related to public key cryptography.
Additionally, the only other pair that made progress did so because they realized that they needed each other's public keys, but even that pair did not know how to then use those shared public keys.
For the remaining eight participant pairs, the post-study interview made it clear that they did not understand how public and private keys were used.

To help address this, a simple explanation of PGP needs to be created that is accessible to the masses.
Several participants indicated that they would prefer these concepts to be displayed in a simple video.
Ideally, whatever form the description takes, it would be be integrated with the tutorials, allowing new users to be introduced to public key cryptography as a natural extension of their usage of Mailvelope.

\subsection{Automatic Email Invites for Recipients}
All participants in the Participant A role were confused about what their friend needed to do in order for their friend to receive encrypted email.
An easy way to address this issue would be to modify PGP-based tools to send an email to recipients for which there isn't an associated public key, asking that individual to install the appropriate software, generate a key pair, and reply with the public key.
This technique was also suggested by Atwater et al.~\cite{atwater2015leading} and our experience corroborates their suggestions.

\subsection{Better Text to Accompany PGP Block}
When first seeing the PGP block generated by Mailvelope, no participant in the Participant B role was clear what they were supposed to do with it.
One participant noted that they thought it was an image that had gotten garbled during email transmission. The one participant that managed to successfully use Mailvelope, M10B, even stated, \textit{``It was like a puzzle, I only got a link to Mailvelope. I then had to go there and explore.''}

To address this issue, PGP-based tools could adopt a technique used by other secure email tools: including plaintext instructions detailing the nature of the encrypted email and how to decrypt it \cite{ruoti2013confused}.
While it is true that instructions on how to decrypt the message are meaningless---as PGP-based tools can only encrypt messages for a recipient who has already established and shared their public key with the sender---there is still room for instructions helping the unexpected recipient of an encrypted email know what to do next.
These instructions could invite the recipient to install the PGP software, generate a key pair, and then send their public key to their friend.
While these instructions aren't foolproof, they are certainly a positive improvement to the current situation.

\section{Conclusion}
We studied Mailvelope, a browser-based PGP secure email tool that integrates tightly with users' existing webmail providers.
In our study of 20 participants, grouped into 10 pairs of participants who attempted to exchange encrypted email, only one pair was able to successfully complete the assigned tasks using Mailvelope.
All other participants were unable to complete the assigned task in the one hour allotted to the study.
Even though a decade has passed since the last formal study of PGP \cite{sheng2006why}, our results show that Johnny has still not gotten any closer to encrypt his email using PGP.

While our results are disheartening, we also discuss several ways that participant experiences and responses indicate how PGP could be improved.
First, integrated tutorials could help first time users with step-by-step instructions.
Second, an approachable description of public key cryptography would give novice individuals a ``fighting chance'' at using PGP.
Third, PGP-based tools would be more effective if, when encountering an unknown recipient, the recipient were to be automatically sent an email telling them what they needed to do receive encrypted email.
Finally, the PGP block itself could be prefixed with instructions that would help non-PGP users understand how to overcome the problem of receiving an encrypted message that they can't decrypt.

\section{Acknowledgements}
We would like to thank Scott Heidbrink, Mark O’Neill, Elham Vaziripour, and Justin Wu for their assistance as study coordinators.

\balance
\bibliographystyle{SIGCHI-Reference-Format}
\bibliography{paper}

\end{document}